\documentclass[10pt,twocolumn]{article}
\usepackage{cite}
\usepackage{graphicx}
\usepackage{pdfpages}
\usepackage{rotating}
\usepackage{placeins}
\usepackage{csvsimple}
\usepackage{listings}
\usepackage{url}
\usepackage{amsmath}
\usepackage{stmaryrd}

\usepackage{siunitx}
\usepackage{booktabs}
\usepackage{xcolor}
\usepackage{setspace} 
\usepackage{verbatim}

\usepackage{MnSymbol}
\usepackage{wasysym}

\newcommand{\modestsections}{
    \let\section=\subsection
        \renewcommand{\thesubsection}{\arabic{subsection}}
    \let\subsection=\subsubsection  
    \let\subsubsection=\paragraph
    \let\paragraph=\subparagraph
    \renewcommand{\subparagraph}[1]{\paragraph{##1}
        \typeout{You've used the subparagraph command, which is the same as
                 the paragraph command since you're using modest sections.}}}

\providecommand{\keywords}[1]{\textbf{\textit{Index terms---}} #1}

\begin{document}


\modestsections

\title{The CATS Hackathon:
Creating and Refining Test Items\\ for Cybersecurity Concept Inventories}


\author{Alan T. Sherman,$^1$ 
Linda Oliva,$^2$\\
Enis Golaszewski,$^1$
Dhananjay Phatak,$^1$
Travis Scheponik,$^1$\\
University of Maryland, Baltimore County (UMBC)\\
Baltimore, Maryland 21250\\
email: \{sherman, oliva, golaszewski, phatak, tschep1\}@umbc.edu\\
\and
Geoffrey L. Herman,$^3$
Dong San Choi,$^3$
Spencer E. Offenberger,$^3$\\
University of Illinois at Urbana-Champaign\\
Champaign, Illinois 61820\\
email: \{glherman, so10\}@illinois.edu\\
\and
Peter Peterson,$^4$
Josiah Dykstra,$^5$
Gregory V. Bard,$^6$
Ankur Chattopadhyay,$^7$\\
Filipo Sharevski,$^8$
Rakesh Verma,$^9$
Ryan Vrecenar$^{10}$
}


\date{{\today}} 

\maketitle

\footnotetext[1]{Cyber Defense Lab, Department of Computer Science and Electrical Engineering}
\footnotetext[2]{Department of Education}
\footnotetext[3]{Computer Science}
\footnotetext[4]{Computer Science Department, University of Minnesota Duluth} 
\footnotetext[5]{U.S. Department of Defense}
\footnotetext[6]{University of Wisconsin, Stout}
\footnotetext[7]{University of Wisconsin, Green Bay}
\footnotetext[8]{DePaul University}
\footnotetext[9]{University of Houston}
\footnotetext[10]{Texas A\&M University}
\setcounter{footnote}{10}

\begin{abstract}

For two days in February 2018, 17 cybersecurity educators and professionals 
from government and industry
met in a ``hackathon'' to refine existing draft multiple-choice test items,
and to create new ones,
for a {\it Cybersecurity Concept Inventory (CCI)}
and {\it Cybersecurity Curriculum Assessment (CCA)} 
being developed as part of the
{\it Cybersecurity Assessment Tools (CATS)} Project.  
We report on the results of the CATS Hackathon, discussing the methods we used to develop test items, 
highlighting the evolution of a sample test item through this process, and
offering suggestions to others who may wish to organize similar hackathons.

Each test item embodies a scenario, question stem, and five answer choices.
During the Hackathon, participants organized into teams to
(1)~Generate new scenarios and question stems, 
(2)~Extend CCI items into CCA items, and generate new answer choices for new scenarios and stems, and 
(3)~Review and refine draft CCA test items. 

The CATS Project provides rigorous evidence-based instruments for assessing and evaluating educational practices;
these instruments can help identify pedagogies and content that are effective in teaching cybersecurity.
The CCI measures how well students understand basic concepts in 
cybersecurity---especially adversarial thinking---after a first course in the field.
The CCA measures how well students understand core concepts after completing a full cybersecurity curriculum.  

\end{abstract}



\section{Introduction}
\label{intro} 

Presently there is no rigorous, research-based method for measuring the quality of cybersecurity instruction. 
Validated assessment tools are needed so that cybersecurity educators have trusted methods 
for discerning whether efforts to improve student preparation are successful.  
The {\it Cybersecurity Assessment Tools (CATS) Project}~\cite{CATS2017} 
provides rigorous evidence-based instruments for assessing and evaluating educational practices.\footnote{\url{
http://www.cisa.umbc.edu/cats/index.html }}
The first instrument is a {\it Cybersecurity Concept Inventory (CCI)} 
that measures how well students understand core concepts in cybersecurity 
(especially adversarial thinking) after a first course in the field. 
The second instrument is a {\it Cybersecurity Curriculum Assessment (CCA)} 
that measures how well students understand the same core concepts after completing a full cybersecurity curriculum
and being ready to enter the workforce as cybersecurity professionals. 
These tools can identify pedagogies and content that are effective in teaching cybersecurity.

In February 2018, we hosted a two-day ``CATS Hackathon'' for 17 cybersecurity educators and professionals 
from across the nation
to generate multiple-choice test items for the CCA, 
and to refine draft items for the CCI and CCA.  
The meeting was a ``hackathon'' in that participants
collaborated on a common task in an informal setting~\cite{Pink11}. 
Over the past couple years, we had developed a bank of about 36 questions for the CCI and about 12 draft questions for the CCA. Participants used these questions as a starting point, extending CCI questions to be CCA questions, refining draft CCA questions, and devising new CCA questions entirely.  
The intimate in-person event facilitated productive interactions among the participants,
infusing fresh ideas into the project, promoting awareness of the tools,
and enhancing the quality of the test items.

We report on this Hackathon, illustrating with an example how a question evolved through three teams 
that conceived new scenarios and question stems, generated answer choices, and 
reviewed draft questions.   
Finally, we document lessons learned from this event.

\section{The CATS Project}
\label{cats} 

Inspired by the {\it Force Concept Inventory (FCI)} of physics by Hestenes, et al.~\cite{Hes92},
we are designing the CCI and CCA to be rigorous assessment tools relevant 
to a wide range of educational contexts. 
 
Unlike the CISSP,\footnote{Certified Information Systems Security Professional (CISSP)
\url{https://www.isc2.org/cissp/default.aspx }}
which is largely informational, our instruments assess {\it conceptual} understanding.  
Like the FCI, our new tests focus only on core concepts to maximize applicability to a variety of curricula 
and thus are intentionally not comprehensive. 
Our tests are intended to measure conceptual understanding, which is a critical, transferable skill.
They do not measure general problem-solving, design, analytical, or interpersonal skills. 
They are intended to compare teaching methods, not individual instructors. 
They are standard tests, broadly applicable to many programs, that can be statistically analyzed 
by established methods~\cite{Ham93,Jorion95,Her2014} 
to produce evidence for the efficacy, or lack thereof, of diverse teaching approaches used in cybersecurity.

Each 50-minute test will comprise approximately 25 multiple-choice questions (MCQs), 
five on each of the following five
core concepts identified through our Delphi Process~\cite{Delphi2016}:\footnote{A {\it Delphi process}
solicits input from a set of subject matter experts to create consensus about contentious decisions,
sharing comments without attributions.}

\begin{enumerate}
\item	Identify vulnerabilities and failures.
\item 	Identify attacks against the CIA\footnote{Confidentiality, Integrity, and Availability (CIA).}
triad and authentication.
\item	Devise a defense.
\item	Identify the security goals.
\item	Identify potential targets and attackers.
\end{enumerate}

Each test item embodies three parts: a {\it scenario}, 
a {\it stem} (question/prompt), and five {\it answer choices} (alternatives). 
Several items may share the same scenario, but each item has a unique stem and answer choices. 
Each stem focuses on one targeted concept, though scenarios may deal with multiple concepts. 
Each stem has exactly one correct (best) choice and four distractors (incorrect answer choices). 
Test items should target the above timeless fundamental concepts, 
not merely factual information that is memorized and recalled.

It is our intent that, for each core concept, the five test items 
encompass a range of difficulty levels.
We recognize, however, that experts tend to be poor judges of the difficulty
of test items, so the actual difficulty of each item will not be reliably known until student testing.


The CATS Team developed draft test items using the following structured process.
Building on the five core concepts identified in our Delphi Process, we created scenarios
and interview prompts, which we used to interview students to uncover their misconceptions~\cite{interviews2017}.
It took significant planning, staff time, and effort to 
carry out, record, transcribe, and analyze these think-aloud interviews. 
Subsequently, in discussions held in a conference room or on Skype, we devised
stems and answer choices.
We based distractors mostly on student misconceptions we uncovered
through the interviews.
Scenarios we developed for these interviews provide rich case studies 
for many learning activities~\cite{exploring2016}.
To test draft questions, we use the PrairieLearn 
System,\footnote{\url{https://prairielearn.readthedocs.io/en/getting-started-docs/ }}
developed at the University of Illinois.

There is evidence that well-crafted MCQs can provide the same type of information 
as do Parson’s problems (open-ended problems).
MCQs are easy to grade and interpret, and there is a robust theory for creating and analyzing them. 
Seventy-six percent of our CCI Delphi experts
agreed or strongly agreed that, 
``A carefully constructed multiple-choice assessment can provide valuable information 
for assessing the quality of instruction in a first course in cybersecurity.''
Other types of assessments (e.g., simulations, hands-on activities, competitions)
also have much to offer but are more complex to
create, maintain, administer, and analyze.

It is essential that these tools have strong usability and validity, 
and that they are implemented widely in diverse settings.  
Throughout, the project benefitted from inputs from a wide variety of experts, 
beginning with our Delphi experts~\cite{Delphi2016}.
We planned the Hackathon to encourage and facilitate experts to collaborate 
on refining existing test items for the CCI and CCA and developing additional test items for the CCA.
The project will continue forward with expert reviews and pilot testing of draft test items.

\section{The Hackathon}
\label{hack} 

To generate new test items for the CCA, the 17 participants organized into several teams, 
each with about three or four members.  
Each team focused on one of the following tasks:
(1)~Generate new scenarios and question stems, 
(2)~Extend CCI items into CCA items, and generate new answer choices for new scenarios and stems, and 
(3)~Review and refine draft CCA test items. 
These substantial tasks kept each team fully engaged throughout the two-day Hackathon.
Each participant chose what team to join, based in part on their skill sets.


The event took place at an off-campus conference center, two days before the
ACM SIGCSE\footnote{ACM Special Interest Group on Computer Science Education (SIGCSE). }
conference in Baltimore.
The experts represented 13 from universities, two from industry, and two from government.  
Participants took the CCI at the beginning of the first day, and 
the CCA at the beginning of the second day.

We now describe each task in more detail.

\subsection{Task 1: Generate New Scenarios and Question Stems} 

These teams started by brainstorming potential scenarios. 
Team members shared their scenarios and developed a priority list 
of ones that needed further development.
Members then refined each scenario by adding details, identifying critical assumptions, 
and drafting 1-4 candidate questions to probe student understanding of the scenario.

The guiding question for this task was, ``Will the new CCA item probe one of the 
identified five core cybersecurity concepts?'' 

We strive to place complexity into the scenarios rather than into the stems.
Doing so helps enable each stem to be as short and clear as possible
and to focus directly on an important concept.
This strategy also reduces the required time for students to complete the test because multiple stems
may share a common scenario.

Participants found it helpful to build on life experiences and to introduce an artifact, such as
a program fragment, log file, protocol, or architectural diagram.

To deemphasize the importance of information knowledge, instead of referring to an object 
(e.g., the SSL protocol), its name or acronym, we described the crucial properties of the object
(e.g., a protocol that encrypts the transferred file, 
using a key established by a key-agreement protocol between sender and receiver.)
To deemphasize vocabulary barriers, we included at the end of each test item definitions of
any terms that students found unfamiliar (e.g., ``masquerade''). 

\subsection{Task 2: Extend CCI Items into CCA Items and Generate New Answer Choices}

These teams focused first on extending existing CCI items 
to have greater technical detail, sophistication, and complexity.
Participants focused on the differences between students 
who have taken only a single course versus students who have taken an entire curriculum in cybersecurity. 

Guiding questions for this task were, ``What do students know?'' 
and ``What misconceptions might students have about this scenario?''

After extending the CCI items into CCA items, 
these teams focused on developing correct answer choices and distractors. 
To ensure that distractors reflected student misconceptions, 
one member of the CATS Team who had previously analyzed student 
misconceptions in cybersecurity contributed his expertise~\cite{interviews2017}. 
These teams exercised leeway to modify scenarios or stems 
as needed to generate compelling and clear correct answers and distractors.

\subsection{Task 3 Review and Refine Draft CCA Test Items}

These teams refined and prioritized draft items 
and made notes about the scenarios, stems, and alternatives for future work.  
The teams first reviewed draft CCA items that the CATS Team had previously created, 
and then reviewed draft CCA items generated by Task~2 teams. 
Members also kept track of how many test items covered each core concept, and 
they estimated the approximate difficulty of each item.

The guiding question for this task was, ``Which scenarios and stems are worthy of inclusion in the CCA?''

These teams focused on quality control, making sure that all wording was precise, concise, and clear. 
One member of the CATS Team experienced in crafting MCQs participated.
Members ensured that each test item stated all critical assumptions. 
Team members answered each draft item and verified that everyone agreed on the correct answer.

Members applied best practices in writing effective multiple-choice questions, including advice offered by
the Vanderbilt Center for Teaching~\cite{Vanderbilt}.  
Each stem should be meaningful by itself.  
The alternatives should be plausible, homogeneous, and non-overlapping.
Each test item should be easy for experts to answer, 
but hard for students with poor or incomplete conceptual understanding.

Many difficulties could be resolved by adding more detail, especially about the assumptions and adversarial model.
Whenever possible, we preferred to insert such details into the scenario rather than into the stem.

\section{An Example: Forensic Analysis of a Network Log File}
\label{example} 

We present a sample CCA test item that originated from Josiah Dykstra at the Hackathon and evolved 
through several discussions and refinements, both during the Hackathon and afterwards by the CATS Team.
Dykstra is a government employee who brought to the Hackathon significant knowledge and experience in
forensics, networks, cybersecurity, and cloud computing.

\begin{figure*}

\noindent {\bf Scenario H2.} {\it Consider the following log of corporate user activity. 
The corporation issues each employee a work PC and a smartphone.}

\medskip

\begin{tabular}{l|l|l|l|l|rr}
Day    & Time     & User    &  Action                    & Device     & \multicolumn{2}{c}{Data Volume [kilobytes]} \\
\hline 
May 21 & 20:22:28 & Bob     & Local login                & Work PC    &       0 UP &       0 DOWN \\
May 21 & 20:23:01 & Bob     & Connection to local server & Work PC    &   6,702 UP & 244,328 DOWN \\
May 21 & 20:25:12 & Bob     & Access to acmeshare.com    & Work PC    & 122,164 UP &   3,456 DOWN \\
May 21 & 20:26:35 & Bob     & USB drive connected        & Work PC    & 122,164 UP &      0 DOWN \\
May 22 & 08:28:12 & Alice   & Connection to remote host  & Work PC    & 122,164 UP &   2,378 DOWN \\
May 22 & 08:32:12 & Charlie & VPN login to network       & Smartphone &   2,490 UP &   4,566 DOWN \\
May 22 & 08:38:55 & Charlie & Access to acmeshare.com    & Smartphone &       0 UP & 125,620 DOWN \\
\end{tabular}

\medskip

\noindent {\it NOTES: \\
(1)~acmeshare.com is a fictional, free file-sharing service.\\
(2)~UP and DOWN data transfer volumes are given from the perspective of the specified device.}

\bigskip \noindent {\bf Question H2-1.} 
{What is the most serious malicious activity possibly suggested by this log?}

\medskip
\begin{tabular}{ll}
    A. & Bob, Alice, and Charlie cooperated to exfiltrate data.\\
    B. & Alice sent corporate secrets to some unspecified remote host.\\
    C. & Bob connected a USB drive and wrote sensitive data to it from his corporate work PC.\\
    D. & Charlie and Bob shared a malicious file via acmeshare.com.\\
    E. & Bob logged in from work at 20:22:28, after the authorized access times.\\
	\end{tabular}

\caption{An example CCA test item that evolved from the Hackathon.}
\label{fig:H2-1}

\end{figure*}

Figure~\ref{fig:H2-1} gives the current polished version of test item H2-1.
It depends on scenario H2, which introduces an artifact that is a network log file of corporate user activity.
Stem H2-1 probes Core Concept~5 (identify targets and attacker) by asking the student to 
identify the most serious malicious activity. 
We suggest that the reader now pause to answer the question.

The CATS Team estimates the difficulty of this test item to be medium.
We consider this test item to be more appropriate for CCA than for CCI because
it requires the student to understand a somewhat technical log file,
however modest the technical aspects may be.

To answer this test item, the student must read and understand the log file
and make inferences about it.  The student must determine who the
adversary or adversaries are and what they have done.  To make these inferences, the student must
demonstrate some technical ability to analyze a log file, common sense, and 
adversarial thinking in a corporate network environment.

To help us keep track of our test items and their status, 
for each test item we assign a line of meta-data summarizing the item's
difficulty, status, core concept, and secondary topic.  
The meta-data for test item H2-1 is:
``Medium, Ready, Identify Targets and Attackers, Log Analysis.''

At the Hackathon, knowing that Dykstra is an expert in forensics, we suggested that
he create a scenario involving forensics.  
Needing more questions involving ``Identify Targets and Attackers,'' we encouraged him
to focus on that concept.  We also suggested that he introduce a technical artifact;
for forensics, the choice of using a log file was natural.

Originally, Dykstra proposed three stems for Scenario~H2,
which we shall call H2-1a, H2-2a, and H2-3a (see Figure~\ref{fig:h2-orig}).
In the ensuing discussions, we settled on only one stem.
H2-3a did not seem to exercise a very important concept, and
H2-1a and H2-2a are overly similar, and the answer to one gives a major hint
of the answer to the other.

\begin{figure*}[t]

\noindent {\bf Question H2-1a.}
Imagine you are an insider stealing corporate secrets. What change would you make in this log to cover your tracks?

\medskip
\begin{tabular}{ll}
A. & Modify all of the data volume entries with random values.\\
B. & Delete the records of login actions.\\
C. & Change all the timestamps to 00:00:00.\\
D. & Erase the action field from all records.\\
E. & Append 500 fake records to the log.\\
\end{tabular}
\bigskip

\noindent {\bf Question H2-2a.}
Which inference can you draw about the attack?

\medskip
\begin{tabular}{ll}
A. & Alice, Bob, and Charlie are colluding in the attack.\\
B. & The attack originated from a remote, external hacker.\\
C. & The firewall is misconfigured.\\
D. & Bob cannot be the attacker.\\
E. & [to be written] \\
\end{tabular}
\bigskip

\noindent {\bf Question H2-3a.}
What other forensic data would implicate the insider(s)?

\medskip
\begin{tabular}{ll}
A. & Network traffic captures.\\
B. & Intrusion detection logs.\\
C. & Firewall logs.\\
D. & Browser history.\\
E. & List of deleted files.\\
\end{tabular}

\caption{The original three stems and their answer choices proposed for Scenario~H2.}
\label{fig:h2-orig}

\end{figure*}

We also modified the stem to focus more directly on the important
targeted concept of identifying what malicious activity took place
and by whom.  As stems should be, Stem H2-1 is a meaningful question
by itself.  

Over multiple meetings, the team spent significant time and effort polishing the test item.  
Much of that effort went into improving the clarity of the item.
It is our experience that many students become confused about various details, including
ones that team members had considered to be clear.  
Small changes in wording can affect how students perceive a test item.
Our instruments should not be 
tests of intelligence or reading comprehension;  
each test item should  challenge a student's conceptual understanding of the targeted concept.

Edits included making the log file more uniform, 
inserting additional information in the log file, 
and clarifying the meaning of data up-loads and down-loads.  
We added clarifying details about the
file-sharing service and who issued the workstations and smartphones.
We also finely edited the wording, for example, replacing the
strong verb ``colluded'' with the softer and less suggestive diction ``cooperated.''
While making such edits may seem simple, 
our experience is that it is difficult and time consuming to construct quality test items.

In case the reader is uncertain, we note that answer choice A is the best alternative 
for each of the above stems.  The sizes of the data flows provide useful clues.

\section{Discussion}
\label{discuss} 

The two-day Hackathon resulted in four promising new CCA test items and useful feedback
on all 36 draft CCI questions and 12 draft CCA questions.
It also increased awareness about our project,
infused new ideas into it, and 
established connections for possible future collaborations.

We learned that our choices for the event---including its size, length, and structure---worked well.
Diversity of the participants, and interactions among them, contributed greatly to the event's success.
Asking the participants to bring some of their favorite questions (e.g., from final exams) is an effective way
to involve everyone from the beginning.

The greatest challenge in running our Hackathon was finding time
in the schedules of busy experts.  
Scheduling the Hackathon physically near and immediately before a major relevant conference 
made it more convenient for participants to attend.
Supporting their travel also helped.

We encountered many challenges in developing quality MCQs:
The process takes a significant amount of time and effort.
Some appealing open-ended questions (e.g., devising or comparing
a design or attack) are difficult to formulate as a MCQ without
depreciating the most attractive aspects of the question.
Often we found it hard to generate more than three appealing distractors.
We strived to make the test items as timeless as possible, but this
goal was challenging to achieve, especially for the more technical CCA.

While most experts liked the majority of our draft questions, some experts disagreed 
with some of our answer choices.  The reason usually involved either
relative weights placed on various considerations or
that the expert made a hidden assumption.  In such cases, we edited the test item
to add details and clarify assumptions.

We are beginning to experiment with and study a new method of generating
distractors:  online crowdsourcing. Using Amazon Mechanical Turk,
we have collected open-ended responses to draft stems.  
Team members analyzed the responses, observing groups of similar incorrect answers, and noting
whether the incorrect responses are consistent with misconceptions that we had expected.  
This method is fast and efficient and benefits from being able to use a specific stem.  
For 25~cents a response, we can easily collect 50--100 responses overnight.
Whereas crowdsourcing is unmoderated with no collaboration and limited control of subjects,
the Hackathon was moderated and facilitated collaboration among carefully selected participants.

It would be helpful to be able to use a suitable integrated test-development
system that supported version control, collaborative test item development,
record keeping, comments, expert review, cognitive interviews, pilot testing, and 
psychometric testing. Unfortunately, we are not aware of any such system.
Instead of building and maintaining our own system, we used a variety of existing
tools, especially ones that support real-time collaboration,
including GitHub, Google Docs (including with the Edity HTML plugin), 
Skype, Survey Monkey, Excel, and PrairieLearn.  
When developing test items, we found it especially helpful to
engage in a remote conference during which the participants could simultaneously
edit a common file.  It is highly desirable, for each test item, to maintain exactly one authentic 
source file, to avoid inevitable errors that will result from copying or converting
test items.  Edity helped us achieve this goal, albeit imperfectly, given that
PrairieLearn inputs test items as HTML files.

As evidenced from feedback submitted via a SurveyMonkey questionnaire, 
the majority of participants found the Hackathon fruitful and 
that it produced valuable products.  
Participants stated that the collaboration with diverse stakeholders was particularly valuable 
to address the diverse and evolving field of cybersecurity education.  
All of the participants indicated that they would be willing to continue contributing
to the development of the instruments and that they would administer pilot versions of the tests.

This year, we are completing development of the draft CCA while beginning validation studies of the CCI.
These validation studies include cognitive interviews, expert review, small-scale pilot testing, 
and large-scale psychometric testing.  
We also plan to carry out several half-day ``mini-hackathons'' associated with various
cybersecurity educational conferences.
We welcome participation in these studies and those to follow for the CCA.

Our experience with the Hackathon demonstrates that this type of 
collaborative workshop is an effective way to generate and improve test items and
to raise awareness about the project.
We hope that the resulting instruments will help identify effective strategies 
for teaching and learning cybersecurity concepts.

\section*{Acknowledgments}
\label{acks} 

We thank all of the Hackathon participants.
This work was supported in part by
the U.S. Department of Defense under CAE-R grants H98230-15-1-0294, 
H98230-15-1-0273, H98230-17-1-0349, and H98230-17-1-0347; and 
by the National Science Foundation 
under SFS grant 1241576
and DGE grant 1820531.

\small
\bibliography{CATShack-bib}
\bibliographystyle{alpha}

\bigskip \noindent
Submitted to {\it IEEE Security \& Privacy} on {\today}.

\end{document}